\documentclass{iopjournal}
\usepackage{lmodern}
\usepackage{bm}
\usepackage{amsmath,amssymb}
\usepackage{CJKutf8}

\begin{document}

\articletype{Paper}

\title{Third Quantization for Order Parameters (II): Field Quantization in Quantum Circuits}

\author{Miao-Miao Yi (\begin{CJK}{UTF8}{gbsn}易淼淼\end{CJK})$^1$, Guo-Jian Qiao (\begin{CJK}{UTF8}{gbsn}乔国健\end{CJK})$^1$, Xin Yue (\begin{CJK}{UTF8}{gbsn}岳鑫\end{CJK})$^2$ and C. P. Sun  (\begin{CJK}{UTF8}{gbsn}孙昌璞\end{CJK})$^{1,*}$}

\affil{$^1$Graduate School of China Academy of Engineering Physics, Beijing 100193, China}

\affil{$^2$Beijing Computational Science Research Center, Beijing 100193, China}

\affil{$^*$Corresponding author.}

\email{suncp@gscaep.ac.cn}

\begin{abstract}
    The quantization of superconducting transmission-line resonators is usually introduced phenomenologically by modeling the resonator as an effective LC circuit and imposing canonical commutation relations on macroscopic variables such as charge and flux. Although this approach has been remarkably successful, it leaves open a fundamental question: why should these macroscopic variables obey quantum commutation relations, and how does such behavior emerge from the spontaneous $U(1)$ symmetry breaking of the superconducting state? In this work, starting from the microscopic pairing Hamiltonian underlying BCS superconductivity, we derive the low-energy effective Hamiltonian of a circuit-QED architecture containing a superconducting transmission line with distributed capacitive and inductive elements. We establish quantitative relations between macroscopic observables, including current and voltage, and the spatially local superconducting phase, as well as the microscopic parameters of the underlying electron-phonon system. On this basis, we extend the \textit{third quantization} for the superconducting order parameter, introduced in Paper (I) of this series for the global phase, to the spatially local case, thereby formulating a macroscopic field quantization. We show that, once the system is restricted to the low-energy excitation subspace, the local superconducting phase necessarily becomes a genuine quantum dynamical variable. In this sense, the quantum behavior of transmission-line resonators and the circuit-QED platforms built from them need not be postulated at the macroscopic level; rather, it follows directly from the \textit{third quantization} for the superconducting order parameter. Our results further indicate that the macroscopic quantum properties of a broad class of superconducting circuit elements, including both transmission-line resonators and superconducting qubits, share the same microscopic origin, thereby establishing \textit{third quantization} as a unified framework for superconducting circuit quantization.
\end{abstract}

\section{Introduction}
The quantization of superconducting transmission lines and their built-in capacitive and inductive elements in circuit QED \cite{Blais2004PRA,Blais2021RMP} underlies a wide range of applications in quantum information processing and quantum computation \cite{PZhang2005PRA,YDWang2005PRB,Review_qubit_computation2001}. Among these components, superconducting transmission-line resonators play a particularly important role. In the standard treatment \cite{Blais2004PRA,Blais2021RMP}, a transmission line is modeled phenomenologically as a distributed LC circuit. One then constructs the corresponding effective Lagrangian \cite{louisell1973book}, chooses the charge $Q(x)$ \cite{Blais2004PRA} or effective flux $\Phi(x)$ \cite{Blais2021RMP} as the canonical field variable, and postulates the associated canonical commutation relations directly at the macroscopic level. This procedure has been highly successful, and many of its predictions have been confirmed experimentally \cite{wallraff2004Nature,majer2007nature}.

Nevertheless, several basic questions remain open. First, the canonical commutation relations of the macroscopic variables are usually taken as the starting point of the theory. It is therefore natural to ask whether these relations should be regarded as new fundamental principles at the macroscopic level, or whether they instead emerge from an intrinsic mechanism already contained in the ordinary second quantization of the underlying many-body system. Closely related to this is a second issue: the connection between this macroscopic quantum phenomenology and the spontaneous $U(1)$ symmetry breaking of the superconducting state has not yet been established in a fully systematic microscopic framework. Third, in the phenomenological approach, macroscopic parameters such as capacitance and inductance are typically introduced from experiment or from effective circuit modeling. It is therefore desirable to derive these quantities quantitatively from microscopic parameters, in particular from the electron-phonon couplings responsible for superconductivity.

A key clue to these questions comes from the quantization of the superconducting order parameter itself. After spontaneous symmetry breaking, the phase of the superconducting order parameter becomes associated with the commutation relation $[\hat N_c, \hat\phi]=\mathrm{i}$, where $\hat N_c$ is the total Cooper-pair-number operator. In our previous work \cite{Qiao2026thirdQuantization}, we showed that, in the thermodynamic limit, this commutation relation can be derived naturally from the second-quantized many-body theory, and we referred to this emergent level of quantization as \textit{third quantization}. From this viewpoint, the macroscopic quantum effects exploited in superconducting qubit devices based on charge, flux, and related circuit degrees of freedom \cite{Moiji_charge_qubit1999,nakamura_charge_qubit1999,Review_qubit_computation2001,Moiji1999} are rooted in the quantization of the order-parameter phase. 

This observation leads naturally to the central question addressed in the present paper: can \textit{third quantization} be extended from the global phase to the spatially local case in a superconducting system described within second quantization? More specifically, can one derive the local field commutation relation $[\hat n_c(x),\hat \phi(x')]=\mathrm{i}\delta(x-x')$, with the Cooper-pair-number density operator $\hat n_c(x)$, and can this commutation relation in turn account for the quantum behavior of macroscopic variables in a superconducting transmission line?

In this paper, we answer these questions affirmatively. Starting from the microscopic pairing Hamiltonian underlying BCS superconductivity, we show that, in the thermodynamic limit, the spatially local form of the commutation relation $[\hat n_c(x),\hat\phi(x')]=\mathrm{i}\delta(x-x')$ emerges naturally in the low-energy excitation space generated by the superconducting ground state. We refer to this local extension of \textit{third quantization} as \textit{macroscopic field quantization}. On this basis, we derive, for the superconducting transmission-line system, the effective inductive and capacitive Hamiltonian terms associated with the LC-circuit description directly from the microscopic theory. We find that the current and flux in the inductive sector are determined by the local phase $\phi(x)$, whereas the voltage in the capacitive sector is determined by the Cooper-pair density $n_c(x)$. The non-commutativity of these macroscopic circuit variables therefore does not arise as a fundamental principle applicable to an arbitrary LC circuit; rather, it is a direct consequence of the quantization of the superconducting order parameter in the low-energy sector of a superconductor.

In addition, we quantitatively establish how macroscopic parameters—including the capacitance density, inductance density, and resonator eigenfrequencies—depend on microscopic parameters of the underlying electron-phonon system. Building on these results, we further construct a systematic correspondence between the basic conjugate variables of \textit{third quantization}, namely the phase and the Cooper-pair number (or density), and the macroscopic variables relevant to different superconducting circuit elements, including superconducting qubits and transmission-line resonators. In this way, the macroscopic quantum phenomena observed in circuit QED can be understood within a unified microscopic framework based on \textit{third quantization}.

\section{Superconducting ground state in real space}\label{sec2}
In this section, starting from attractive Hubbard model, we determine the variational ground state of a superconducting system. The resulting real-space ground state will serve as the starting point for the discussion of the quantization of the spatially local order parameter phase in the next section. For simplicity, throughout this paper we restrict ourselves to the zero-temperature case. The attractive Hubbard model \cite{Hubbard1981PRB,Hubbard1990RMP,zhu2016bogoliubov} considered here is
\begin{equation}
    \hat H=-t\sum_{\langle \bm i\bm j\rangle,\sigma}\hat c_{\bm i,\sigma}^\dagger \hat c_{\bm j,\sigma}-\sum_{\bm i,\sigma}(eV_{\bm i}+\mu)\hat n_{\bm i,\sigma}-U\sum_{\bm i}\hat n_{\bm i,\uparrow}\hat n_{\bm i,\downarrow},\label{Hubbard_model}
\end{equation}
where $\bm i=(i_x,i_y,i_z)$ labels real-space lattice sites, $\langle \bm i\bm j \rangle$ denotes nearest-neighbor pairs, $t$ is the nearest-neighbor hopping amplitude, $-eV_{\bm i}$ is the external potential at site $\bm i$, $\mu$ is the chemical potential, $\hat n_{\bm i,\sigma}:=\hat c^\dagger_{\bm i,\sigma}\hat c_{\bm i,\sigma}$ is the number operator for spin $\sigma$ at site $\bm i$, and $U>0$ denotes the onsite interaction strength. Notably, this Hamiltonian is invariant under the global $U(1)$ transformation $\hat c_{\bm i,\sigma}\rightarrow \exp(\mathrm{i}\hat N\theta)\hat c_{\bm i,\sigma}\exp(-\mathrm{i}\hat N\theta)=\exp(\mathrm{i}\theta)\hat c_{\bm i,\sigma}$, with
$\hat N:=\sum_{\bm i,\sigma}\hat c_{\bm i,\sigma}^\dagger\hat c_{\bm i,\sigma}$, and therefore possesses a global $U(1)$ symmetry.

To describe electron pairing, we adopt the real-space Gaussian variational ground state \cite{THOULESS1960,TaoShiPRD2018}
\begin{equation}
    |GS[\Psi]\rangle=\mathcal{N}\exp(\sum_{\bm i\bm j}|\Psi_{\bm i\bm j}|e^{\mathrm{i}\phi_{\bm i\bm j}}\hat c_{\bm i,\uparrow}^\dagger \hat c_{\bm j,\downarrow}^\dagger)|\bm 0\rangle,\label{viriational_ground_state}
\end{equation}
where $\mathcal{N}$ is a normalization constant, $|\bm 0\rangle:=|0_1,\cdots,0_N\rangle$ is the vacuum state in the system, and $\Psi_{\bm i\bm j}=|\Psi_{\bm i\bm j}|\exp(\mathrm{i}\phi_{\bm i\bm j})$ characterizes the real-space pairing field. Minimization of the zero-temperature free energy functional $F[\Psi]:=\langle GS[\Psi]|\hat H|GS[\Psi]\rangle$ yields
\begin{equation}
    \phi_{\bm i\bm j}=\phi,\label{phi}
\end{equation}
showing that all pairing components share a common global phase. At the same time, the amplitude $|\Psi_{\bm i\bm j}|$ satisfy the corresponding self-consistency equation
\begin{equation}
    -t\sum_{\hat\delta}(|\Psi_{\bm i+\hat\delta,\bm j}|+|\Psi_{\bm i,\bm j+\hat\delta}|)-(eV_{\bm i}+\mu+U n_{\bm i,\downarrow})|\Psi_{\bm i\bm j}|-(eV_{\bm j}+\mu+U n_{\bm j,\uparrow})|\Psi_{\bm i\bm j}|+\tilde\Delta_{\bm i}\delta_{\bm i\bm j}-\sum_m |\Psi_{\bm i\bm m}|\tilde\Delta_{\bm m}|\Psi_{\bm m\bm j}|=0,\label{self_consistency_main_text}
\end{equation}
where $n_{\bm i,\sigma}:=\langle GS[\Psi]|\hat c_{\bm i,\sigma}^\dagger\hat c_{\bm i,\sigma}|GS[\Psi]\rangle$, and $\hat\delta$ runs over the six near-neighbor directions, $\pm\hat x$, $\pm\hat y$ and $\pm\hat z$, so that $\bm i+\hat\delta$ denotes a nearest-neighbor site of $\bm i$. In particular, this real-space variational state can also reproduce the conventional BCS state in the spatially uniform limit $V_{\bm i}=V$. The detailed derivation and discussion are presented in the Appendix \ref{Appendix_variant}.

In this variational description, the superconducting order parameter is defined as
\begin{equation}
    \Delta_{\bm i}:=-U\langle GS[\Psi]| \hat c_{\bm i,\downarrow}\hat c_{\bm i,\uparrow}|GS[\Psi]\rangle=\tilde\Delta_{\bm i}\exp({\mathrm{i}\phi}),\label{order_paramter}
\end{equation}
with
\begin{equation}
    \tilde\Delta_{\bm i}:=-U|\langle GS[\Psi]|\hat c_{\bm i,\downarrow}\hat c_{\bm i,\uparrow}|GS[\Psi]\rangle|=-U[\tilde\Psi(1+\tilde\Psi^2)^{-1}]_{\bm i\bm i},
\end{equation}
where $\tilde\Psi:=[|\Psi_{\bm i\bm j}|]$ denotes the real matrix formed by the amplitudes $|\Psi_{\bm i\bm j}|$. Eq.~(\ref{order_paramter}) shows that the superconducting order parameter inherits the same unified macroscopic phase $\phi$.  A nonzero $\Delta_{\bm i}$ therefore signals the formation of a paired condensate and, equivalently, the spontaneous breaking of the global $U(1)$ symmetry at the level of the ground state, i.e., $\exp(\mathrm{i}\hat N\theta)|GS[\Psi]\rangle\neq|GS[\Psi]\rangle$, $\forall\theta$.

Notably, this spontaneous symmetry breaking implies that the ground state is no longer unique. Instead, the system exhibits an infinitely degenerate ground-state space labeled by the macroscopic phase $\phi$, whose elements are
\begin{equation}
    |GS[\Psi(\phi)]\rangle=\mathcal{N}\exp(\sum_{\bm i\bm j}|\Psi_{\bm i\bm j}|e^{\mathrm{i}\phi}\hat c_{\bm i,\uparrow}^\dagger\hat c_{\bm j,\downarrow}^\dagger)|\bm 0\rangle=\exp(\mathrm{i}\hat N\phi/2)|GS[\Psi(0)]\rangle.\label{ground_state}
\end{equation}
This means that, when the system is in a superposition of ground state associated with different values of $\phi$, it no longer possesses a definite global phase. The phase $\phi$ should therefore be treated as a quantum variable and promoted to an operator $\hat\phi$. This is precisely what was referred to in our previous paper as \textit{third quantization}, namely the quantization of the superconducting order parameter phase degree of freedom. In the present paper, one can further demonstrate that, even when the discussion is extended to inhomogeneous systems, a Hermitian operator $\hat\phi$ can still be defined in the thermodynamic limit, and it forms a canonically conjugate pair with the total cooper-pair-number operator $\hat N_c:=\hat N/2$, i.e., the commutation relation $[\hat N_c,\hat\phi]=\mathrm{i}$. The detailed derivation is presented in the Appendix \ref{appendix_phase}.

\section{Third Quantization for the spatially local order parameter}\label{section_local_phase}

In this section, we extend the \textit{third quantization} to the spatially local case, and refer to this extension as \textit{macroscopic field quantization}. Specifically, when the system is in low-energy excitations relative to the ground state, the phase is no longer confined to the global variable $\phi$, but develops into a slowly varying site-dependent quantity $\phi_{i_x}$. These local phase can be viewed as collective modes associated with the Nambu-Goldstone mode \cite{Goldstone1962RP}. In the thermodynamic limit and the continuum limit $\phi_{i_x}\rightarrow\phi(x)$, these local phase degrees of freedom can likewise be quantized and satisfy the field commutation relation $[\hat n_c(x),\phi(x')]=\mathrm{i}\delta(x-x')$. The \textit{macroscopic field quantization} provides the foundation for the quantization of superconducting transmission-line resonators in quantum circuits.

\begin{figure}
    \centering
    \includegraphics[width=0.5\linewidth]{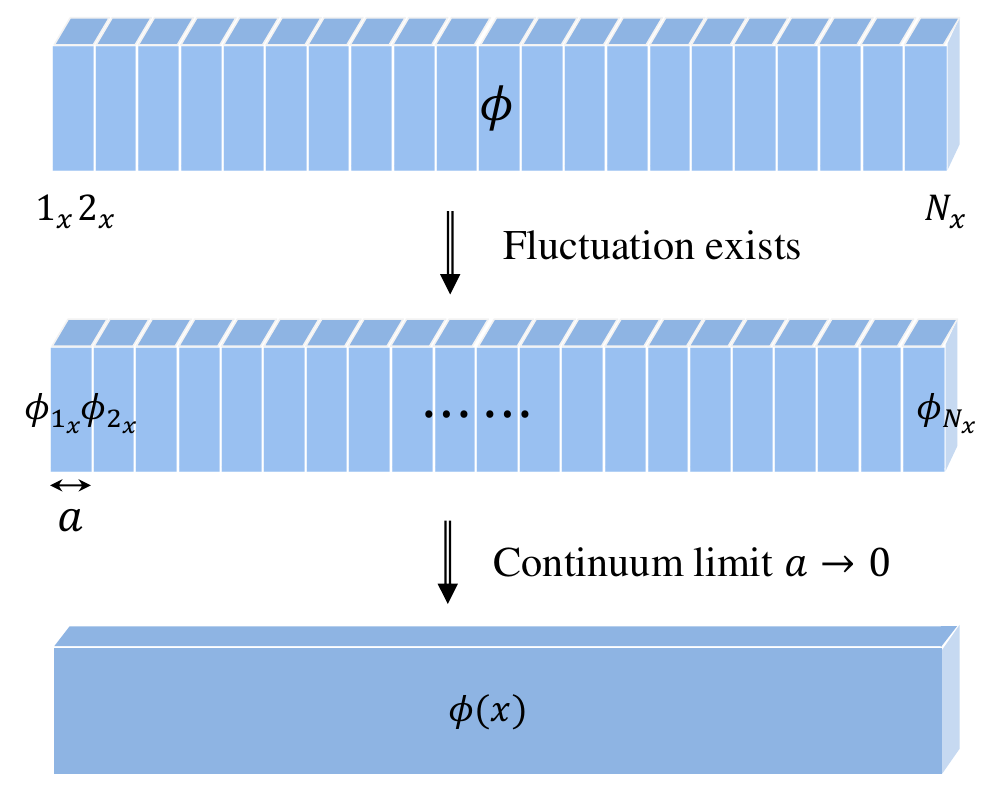}
    \caption{Starting from a uniform global phase $\phi$
 established in the ground state, fluctuations $\phi_{i_x}$ emerge at each lattice site, followed by the continuum limit to obtain the continuous phase field $\phi(x)$.}
    \label{fig_phase_sketch}
\end{figure}

As illustrated in Fig.~\ref{fig_phase_sketch}, the order parameter phase acquires slowly varying site-dependent fluctuation, i.e., $\phi_{i_x}=\phi+\delta\phi_{i_x}$, where $i_x$ labels sites along the $x$-direction. In this case, the discussion should then be enlarged from the uniform-phase ground-state space $\{|GS[\Psi(\phi)]\rangle\}$ to a low-energy phase-excitation space. For a given set of local phases $\{\phi_{i_x}\}$, the corresponding low energy state is
\begin{equation}
    |\psi\{\phi_{i_x}\}\rangle=\exp(\mathrm{i}\sum_{i_x}\phi_{i_x}\hat N_{i_x}/2)|GS[\Psi(0)]\rangle,\label{low_energy_excited_state}
\end{equation}
where $\hat N_{i_x}:=\sum_{i_y,i_z,\sigma} \hat c_{\bm i,\sigma}^\dagger \hat c_{\bm i,\sigma}$ is the electron-number operator on the cross section at $i_x$. When all local phases satisfy $\phi_{i_x}=\phi$, the state reduces to the ground state $|GS[\Psi(\phi)]\rangle$ with a uniform phase $\phi$. 

We now clarify the relation between the local order parameter phase $\phi_{i_x}$ and the Nambu-Goldstone mode. When $\phi_{i_x}$ varies slowly from site to site, i.e., $\delta\phi_{i_x}\ll1$, the excitation energy relative to the ground state is
\begin{align}
    \Delta E=&\,\langle\psi\{\phi_{i_x}\}|\hat H|\psi\{\phi_{i_x}\}\rangle-\langle GS|\hat H|GS\rangle\\
    \simeq&\,\frac{t}{8}\sum_{\langle\bm i\bm j\rangle}\rho_{\bm i\bm j}(\phi_{j_x}-\phi_{i_x})^2\ll1,
\end{align}
where $\rho_{\bm i,\bm j}:=\sum_\sigma\langle\hat c^\dagger_{\bm i,\sigma}\hat c_{\bm j,\sigma}\rangle_0$, with $\langle\cdot\rangle_0:=\langle GS|\cdot|GS\rangle$. This shows that a slowly varying phase profile carries only a small energy cost proportional to the square of its gradient $\nabla\phi\sim(\phi_{j_x}-\phi_{i_x})$, and therefore corresponds to a gapless long-wavelength low-energy excitation above the ground state. According to the Nambu-Goldstone theorem \cite{Goldstone1962RP,Nambu}: \textit{whenever the Hamiltonian of a system possesses a continuous global symmetry that is spontaneously broken in the thermodynamic limit, there must exist gapless low-energy collective excited modes (Nambu-Goldstone modes)}. In a superconducting system, the local phase $\phi_{i_x}$ can be viewed as the concrete manifestation of the corresponding Nambu-Goldstone mode arising from spontaneous breaking of the global $U(1)$ symmetry.

When the system is in a superposition of low-energy states $|\psi\{\phi_{i_x}\}\rangle$ associated with different local phase configurations $\{\phi_{i_x}\}$, the spatially local phase degrees of freedom acquire quantum uncertainty. It is therefore necessary to promote $\{\phi_{i_x}\}$ to operator $\{\hat\phi_{i_x}\}$, i.e., to a local form of \textit{third quantization}. This treatment can be understood as the quantization of the Nambu-Goldstone mode. As in the discussion of Appendix \ref{appendix_phase}, we first show that the low-energy states associated with different local phase configurations become mutually orthogonal in the thermodynamic limit, so that they can serve as eigen-states of the Hermitian operators $\{\hat\phi_{i_x}\}$.

For the one-dimensional superconductor shown in Fig.~\ref{fig_resonator_sketch}, the system is uniform in the transverse $y-$ and $z-$ directions. The amplitude $|\Psi_{\bm i\bm j}|$ is therefore translationally invariant in the cross-sectional plane. We therefore Fourier transform $\hat c_{\bm i,\sigma}$ along the $y-$ and $z-$ directions,
\begin{equation}
    \hat c_{i_x,\bm k_\perp,\sigma}=\frac{1}{\sqrt{N_yN_z}}\sum_{i_y,i_z}e^{-\mathrm{i}(k_yi_y+k_zi_z)}\hat c_{\bm i,\sigma}.
\end{equation}
Hence the low-energy state in Eq.~(\ref{low_energy_excited_state}) reads
\begin{equation}
    |\psi\{\phi_{i_x}\}\rangle=\prod_{k_\perp}^{k_M}\exp(\mathrm{i}\sum_{i_x,\sigma}\phi_{i_x}\hat n_{i_x,\bm k_\perp,\sigma}/2)|GS_{\bm k_\perp}\rangle,
\end{equation}
where $\hat n_{i_x,\bm k_\perp,\sigma}:=\hat c^\dagger_{i_x,\bm k_\perp,\sigma}\hat c_{i_x,\bm k_\perp,\sigma}$, and $M$ is the number of transverse modes, equal to the number $N_yN_z$ of lattice sites on the cross section. Here,
\begin{equation}
    |GS_{\bm k_\perp}\rangle:=\mathcal{N}\exp(\sum_{i_x,j_x}|\Psi_{i_x,j_x,\bm k_{\perp}}|\hat c^\dagger_{i_x,\bm k_\perp,\uparrow}\hat c^\dagger_{j_x,-\bm k_\perp,\downarrow})|\bm 0\rangle,
\end{equation}
where
\begin{equation}
    |\Psi_{i_x,j_x,\bm k_{\perp}}|:=\frac{1}{\sqrt{N_yN_z}}\sum_{i_y-i_z,j_y-j_z}e^{-\mathrm{i}[k_y(i_y-j_y)+k_z(i_x-j_z)]}|\Psi_{\bm i\bm j}|.
\end{equation}
The Cauchy-Schwarz inequality gives
\begin{equation}
    |\langle GS_{\bm k_\perp}|\exp[\mathrm{i}\sum_{i_x,\sigma}(\phi_{i_x}-\phi_{i_x}')\hat n_{i_x,\bm k_\perp,\sigma}/2]|GS_{\bm k_\perp}\rangle|^2\leq 1,
\end{equation}
in general, equality holds only if $\phi_{i_x}=\phi_{i_x'}$, $\forall\, i_x$. Therefore, when the number of lattice sites on the cross section approaches the thermodynamic limit $N_yN_z\rightarrow\infty$ (i.e., $M\rightarrow\infty$), the low-energy states associated with different local phase configurations become naturally orthogonal, i.e., 
\begin{align}
    |\langle \psi\{\phi_{i_x}\}|\psi\{\phi_{i_x}'\}\rangle|&\,=\prod_{k_\perp}^{k_M}|\langle GS_{\bm k_\perp}|\exp[\mathrm{i}\sum_{i_x,\sigma}(\phi_{i_x}-\phi_{i_x}')\hat n_{i_x,\bm k_\perp,\sigma}/2]|GS_{\bm k_\perp}\rangle|\\ &\,\rightarrow\delta_{\phi_{1_x}\phi_{1_x}'}\cdots\delta_{\phi_{N_x}\phi_{N_x}'}.\notag
\end{align}

In analogy with Appendix \ref{appendix_phase}, we define the continuously normalized eigen-states of $\{\hat\phi_{i_x}\}$ as
\begin{equation}
    |\{\phi_{i_x}\}\rangle:=(\prod_{i_x}\lim_{\delta\phi_{i_x}\rightarrow0}\frac{1}{\sqrt{\delta\phi_{i_x}}})|\psi\{\phi_{i_x}\}\rangle,\label{local_phase_eigenstate}
\end{equation}
which satisfy $\langle\{\phi_{i_x}\}|\{\phi_{i_x}'\}\rangle=\prod_{i_x}\delta(\phi_{i_x}-\phi'_{i_x})$. The local phase operators can then be defined in the low-energy excitation space as
\begin{equation}
    \hat\phi_{i_x}=\int^{2\pi}_0 \phi_{i_x}\prod_{i_x}\mathrm{d}\phi_{i_x}|\{\phi_{i_x}\}\rangle\langle\{\phi_{i_x}\}|.\label{local_phi_operator}
\end{equation}

It follows from Eqs.~(\ref{low_energy_excited_state}) and (\ref{local_phase_eigenstate}) that
\begin{equation}
    -2\mathrm{i}\partial_{\phi_{i_x}}|\{\phi_{i_x}\}\rangle=\hat N_{i_x}|\{\phi_{i_x}\}\rangle,
\end{equation}
For an arbitrary state $|\varphi\rangle\in\{|\{\phi_{i_x}\}\rangle\}$ in the low-energy excitation space,
\begin{equation}
    \langle\{\phi_{i_x}\}|\hat N_{i_x}|\varphi\rangle=2\mathrm{i}\partial_{\phi_{i_x}}f(\{\phi_{i_x}\}),
\end{equation}
where $f(\{\phi_{i_x}\}):=\langle\{\phi_{i_x}\}|\varphi\rangle$. This shows that the cooper-pair-number operator $\hat N_{c,i_x}:=\hat N_{i_x}/2=\mathrm{i}\partial_{\phi_{i_x}}$ in the  $|\{\phi_{i_x}\}\rangle$ representation, therefore,
\begin{equation}
    [\hat N_{c,i_x},\hat\phi_{i_x'}]=\mathrm{i}\delta_{i_x,i_x'}.\label{commutation_relation_local}
\end{equation}

Finally, introducing the Cooper-pair line density $\hat n_c(x):=\hat N_{c,i_x}/a$ and identifying $\hat\phi(x):=\phi_{i_x}$, with $a$ the lattice constant, the local commutation relation (\ref{commutation_relation_local}) in the continuum limit $a\rightarrow0$ is written as a canonical commutation relation between field operators,
\begin{equation}
    [\hat n_c(x),\hat\phi(x')]=\mathrm{i}\delta(x-x').
\end{equation}
This canonically conjugate relation emerges naturally from spontaneous symmetry breaking in a superconductor and provides a more fundamental microscopic basis for the quantization of one-dimensional superconducting transmission-line resonators in quantum circuits.

\section{Circuit-Quantization via Order Parameter Quantization}
In this section, we will derive the capacitive-inductive Hamiltonian of a superconducting transmission-line circuit from the BCS theory of superconductivity. The resulting Hamiltonian can be expressed in terms of the local order parameter phase $\phi(x)$ and the cooper-pair-number density $n_c(x)$. We further establish quantitative relations between the macroscopic variables, and both the order parameter phase and the microscopic parameters of the superconductor. In this way, the macroscopic quantum behavior of superconducting transmission-line resonators, especially the non-commutativity of macroscopic variables, is not an independent postulation in quantum mechanics, but instead emerges naturally in the low-energy effective theory from the \textit{third quantization}.
\begin{figure}
    \centering
    \includegraphics[width=0.6\linewidth]{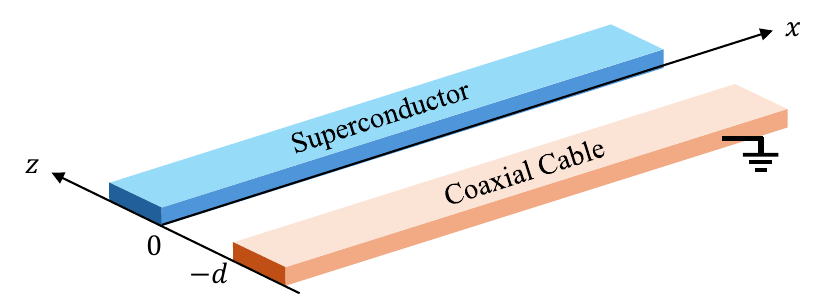}
    \caption{Simplified schematic of superconducting transmission line.}
    \label{fig_resonator_sketch}
\end{figure}
Fig.~\ref{fig_resonator_sketch} illustrates a simplified structure of a one-dimensional superconducting transmission-line resonator, which can be coupled to a superconducting qubit through the voltage relative to a coaxial cable. We consider a superconductor in the presence of electromagnetic fields, and assume that the system is uniform across the transverse directions, i.e., along $y-$ and $z-$ directions, with the number of particles on each cross section taken in the thermodynamic limit. For convenience, we consider the Hamiltonian in the continuous-space mean-field form \cite{de2018superconductivity,zhu2016bogoliubov,tinkham2004introduction}
\begin{align}
    \hat {\mathcal{H}}=&\,\int\mathrm{d}^3\bm x\sum_{\sigma}\hat\psi_{\sigma}^\dagger(\bm x)[\frac{1}{2m}(-\mathrm{i}\hbar\nabla+e\bm A(\bm x))^2-eV(\bm x)-\notag\\
    &\,-\mu]\hat\psi_{\sigma}(\bm x)+\int\mathrm{d}^3\bm x\Delta(\bm x)\psi_{\uparrow}^\dagger(\bm x)\hat\psi_{\downarrow}^\dagger(\bm x)+\mathrm{h.c.},\label{Hamiltonian_LC_microscopic}
\end{align}
where $e$ is the electron charge and  $\hat\psi_{\sigma}(\bm x)$ is the field operator for spin $\sigma$. The vector and scalar potentials of the electromagnetic field are denoted by $\bm A(\bm x)$ and $V(\bm x)$. Here, we take the Coulomb gauge, $\nabla\cdot\bm A=0$, under which $V(\bm x)$ formally satisfies the Possion equation
\begin{equation}
    \nabla^2V(\bm x)=e n(\bm x)/\epsilon,\qquad z>-d,\label{Possion}
\end{equation}
where $n(\bm x):=\sum_{\sigma}\langle\hat\psi^\dagger_{\sigma}(\bm x)\hat\psi_{\sigma}(\bm x)\rangle$ is the electron density at position $\bm x$, $\epsilon$ is the dielectric constant, and $z>-d$ specifies the region in the structure shown in Fig.~\ref{fig_resonator_sketch} where the Poisson equation applies, $d$ is the distance between resonator and the coaxial cable. The order parameter in the continuum description is defined as
\begin{equation}
    \Delta(\bm x)=|\Delta(\bm x)|e^{\mathrm{i}\phi(\bm x)}:=-U\langle\hat\psi_{\downarrow}(\bm x)\hat\psi_{\uparrow}(\bm x)\rangle,
\end{equation}
which corresponds to Eq.~(\ref{order_paramter}) in the discrete setting. Here $\phi(\bm x)$ is the local order parameter phase. Since the system is uniform along $y-$ and $z-$ directions, we write $\phi(\bm x)$ as $\phi(x)$ below.

Next, we will derive the LC Hamiltonian from Eq.~(\ref{Hamiltonian_LC_microscopic}). As shown in Sec.~\ref{sec2}, the appearance of a local phase $\phi(x)$ corresponds to the system being in the space $\{|\phi(x)\rangle\}$ spanned by low-energy phase-excited states. The effective Hamiltonian of superconductor projected onto this low-energy space is therefore
\begin{align}
    \hat H_{\mathrm{eff}}=&\,(\int\mathcal{D}\phi |\phi(x)\rangle\langle\phi(x)|)\hat{\mathcal{H}}(\int\mathcal{D}\phi' |\phi'(x)\rangle\langle\phi'(x)|)\notag\\
    \simeq&\,\int\mathcal{D}\phi |\phi(x)\rangle\langle\phi(x)|H_{\mathrm{eff}},\label{effective_Hamiltonian}
\end{align}
where only the lowest-order contributions are retained. Here, $|\phi(x)\rangle:=\lim_{a\rightarrow0}|\{\phi_{i_x}\}\rangle$ is the continuum-limit low-energy excited state as the lattice constant $a\rightarrow0$, and $\mathcal{D}\phi:=\lim_{a\rightarrow0}\prod_{i_x}\mathrm{d}\phi_{i_x}$. The effective Hamiltonian $H_{\mathrm{eff}}:=H_0+H_1+H_2$ reads
\begin{align}
    H_0=&\,\langle\sum_{\sigma}\int\mathrm{d}^3\bm x\hat\psi_{\sigma}^\dagger(\bm x)[-\frac{\hbar^2}{2m}\nabla^2-\mu]\hat\psi_{\sigma}(\bm x)+|\Delta(\bm x)|\hat\psi^\dagger_{\uparrow}(\bm x)\hat\psi_{\downarrow}^\dagger(\bm x)+\mathrm{h.c.}\rangle_0,\label{H_0}\\
    H_1=&\,\frac{\hbar}{2e}\int\mathrm{d}^3\bm x\langle\bm {\hat J}_p(\bm x)\rangle_0\cdot\nabla\varphi(\bm x)+\frac{\hbar^2}{8m}\int\mathrm{d}^3\bm x n(\bm x)(\nabla\varphi(\bm x))^2\notag\\
    \simeq&\,\frac{\hbar^2}{8m}\int\mathrm{d}^3\bm x n(\bm x)(\nabla\varphi(\bm x))^2,\label{H_1}\\
    H_2=&\,-e\int\mathrm{d}^3\bm x n(\bm x) V(\bm x).\label{H_3}
\end{align}
where $\langle\cdot\rangle_0:=\langle GS|\cdot|GS\rangle$, and $|GS\rangle$ is the ground state (\ref{ground_state}) of the system at $\bm A=\bm0$ (i.e., the system described by Eq.~(\ref{Hubbard_model})). We further assume that the magnetic field vanishes identically inside the superconductor, i.e., $\nabla\times\bm A=\bm0$, hence we define the gauge-invariant phase
\begin{equation}
    \varphi(\bm x):=\phi(\bm x)-\frac{2e}{\hbar}\int^{\bm x}_0\bm A\cdot\mathrm{d}\bm l.\label{effective_phase}
\end{equation}
In addition,
\begin{equation}
    \hat{\bm J}_p(\bm x):=\frac{\mathrm{i}\hbar e}{2m}\sum_{\sigma}(\hat\psi^\dagger_{\sigma}(\bm x)\nabla\hat\psi_{\sigma}(\bm x)-\hat\psi_{\sigma}(\bm x)\nabla\hat\psi^\dagger_{\sigma}(\bm x))
\end{equation}
is the paramagnetic current operator. Under weak electromagnetic fields ($A,\,V\ll1$), one has $\langle\hat{\bm J}_p(\bm x)\rangle_0\simeq0$ \cite{de2018superconductivity}. As we show below, $H_1$ and $H_2$ reproduce the effective inductive and capacitive energies of a quantum circuit, respectively.

We first consider the effective inductive structure generated by the vector-potential term. Because the system is uniform over each cross section perpendicular to the $x$ axis, the Hamiltonian (\ref{H_1}) is rewritten as
\begin{equation}
    H_1= \frac{n_x\hbar^2}{8m}\int\mathrm{d}x(\partial_x\varphi)^2+\frac{\hbar^2}{8m}\int\mathrm{d}x\delta n(x)(\partial_x\varphi)^2,\label{inductance0}
\end{equation}
where $n_x:=\int\mathrm{d}x n(x)/L_x$ is the average number density along the $x$ axis, $n(x)$ denotes the electron number per unit length, and $\delta n(x):=n(x)-n_x$ describes the density fluctuation.

Moreover, the current through the $y$, $z$ cross section is
\begin{equation}
    I(x)=\langle\bm {\hat J}\rangle\cdot \bm{S}_{y,z}\simeq\frac{e\hbar n_x}{2m}\partial_x\varphi,\label{current}
\end{equation}
with $\langle\cdot\rangle:=\langle\phi(x)|\cdots|\phi(x)\rangle$. Here $\hat{\bm J}$ denotes the total current density operator, which satisfies $\nabla\cdot\hat{\bm J}+(-e\hat n(\bm x))=0$. The above approximation follows from $\langle\hat{\bm J}_p(\bm x)\rangle_0\simeq0$ under weak electromagnetic fields. Substituting Eq.(\ref{current}) into Eq.(\ref{inductance0}), and retaining only the leading term in Eq.~(\ref{inductance0}) under the assumption that $n(x)$ varies slowly in space, one obtains
\begin{equation}
    H_1\simeq\int\mathrm{d}x\frac{1}{2}l I^2(x)=\int\mathrm{d}x\frac{1}{2l}(\frac{\hbar}{2e}\partial_x\varphi)^2.\label{H_ind}
\end{equation}
The effective inductance per unit length $l$ is therefore
\begin{equation}
    l=\frac{m}{n_x e^2}.\label{inductance_per_unit}
\end{equation}
This establishes a direct relation between the effective inductance per unit length and the microscopic parameters of the superconductor.

Next, we consider the effective capacitive structure generated by the scalar-potential term. For the structure shown in Fig.~\ref{fig_resonator_sketch}, we impose the boundary condition $V(x,y,z=-d)=0$. Combined with the Poisson Equation (\ref{Possion}), the boundary condition yields the unique solution
\begin{equation}
    V(\bm x)=-e\int\mathrm{d}^3\bm x' G(\bm x,\bm x')n(\bm x'),\label{scalar_potential}
\end{equation}
 Therefore,
\begin{equation}
    H_2=e^2\int\mathrm{d}^3\bm x\mathrm{d}^3\bm x' n(\bm x)G(\bm x,\bm x')n(\bm x'),\label{H_capa}
\end{equation}
where the Green's function is
\begin{equation}
    G(\bm x,\bm x')=\frac{1}{4\pi\epsilon}(\frac{1}{|\bm x-\bm x'|}-\frac{1}{|\bm x-\bm x''|}),
\end{equation}
with $\bm x''=(x',y',-2d-z')$ and $\bm x'=(x',y',z')$.

We further approximate the superconductor as an ideal conductor, with $\epsilon\rightarrow\infty$ in the interior and charge confined to the surface $z=0$. Then $n(\bm x)\simeq\delta(z)n(x)/L_y$, where $L_y$ is the thickness along the $y-$ direction. Eq.~(\ref{H_capa}) then becomes
\begin{equation}
    H_1=e^2\frac{1}{2\pi L_y}\int\mathrm{d}k_x n(k_x)n(-k_x) G(k_x),
\end{equation}
where
\begin{equation}
    n(k_x):=\int\mathrm{d}x\exp(-\mathrm{i}k_xx) n(x),
\end{equation}
and
\begin{equation}
    G(k_x)=\frac{1}{2\epsilon|k_x|}[1-\exp(-2|k_x|d)].
\end{equation}
Under the assumption that $n(x)$ varies slowly in space, with $\partial_x n(x)\lesssim[n(x)/L_x]_{\max}$, the nonzero weight of $n(k_x)$ is concentrated near $k_x\ll 1/L_x$. It then follows that
\begin{equation}
    H_2\simeq\frac{e^2}{2\pi L_y}\int\mathrm{d}k_x n(k_x)n(-k_x)G(k_x=0)=\int\mathrm{d}x\frac{1}{2c}(en(x))^2.\label{H_cap}
\end{equation}
Under the same slowly varying condition, the voltage of the transmission-line resonator relative to the coaxial cable is approximated as $V(x):=V(x,y,z=0)-V(x,y,z=-d)\simeq -e n(x)/c$. The effective capacitance per unit length $c$ is given by
\begin{equation}
    c=\frac{L_y\epsilon}{2d},\label{capacitance_per_unit}
\end{equation}
so that the scalar-potential term reproduces the capacitive energy a effective quantum LC circuit.

Substituting Eqs.~(\ref{H_ind}) and (\ref{H_cap}) into the effective Hamiltonian (\ref{effective_Hamiltonian}), and using the definition (\ref{local_phi_operator}), one promotes the local phase $\varphi(x)$ (equivalently $\phi(x)$) and the electron-number density $n(x)$ to operators. This is precisely how \textit{third quantization} is realized at the level of the macroscopic circuit. Under the condition that the external fields $\bm A$, $U$ and the density $n(x)$ vary slowly in space, the resulting low-energy effective Hamiltonian is
\begin{equation}
    \hat H_{\mathrm{eff}}\simeq\int\mathrm{d}x\frac{1}{2l}(\frac{\hbar}{2e}\partial_x\hat\varphi)^2+\int\mathrm{d}x\frac{1}{2c}(e\hat n(x))^2,\label{effective_Hamiltonian_quantum}
\end{equation}
Here, $H_0$ contains neither $\phi(x)$ nor $n(x)$. It therefore contributes only an additive constant term, and is omitted in what follows.

According to Eqs.~(\ref{commutation_relation_local}) and (\ref{effective_phase}), $\hat\varphi(x)$ and $\hat n(x)$ satisfy
\begin{equation}
    [\hat n(x),\hat\varphi(x')]=2\mathrm{i}\delta(x-x').
\end{equation}
On this basis, we further perform the canonical mode decomposition of the Hamiltonian \eqref{effective_Hamiltonian_quantum}. The resulting Hamiltonian is
\begin{equation}
    \hat H_{\mathrm{eff}}=\sum_j\hbar\omega_j(\hat a^\dagger_j\hat a_j+\frac{1}{2}),
\end{equation}
where $\omega_j=j\omega_1$ are the eigen-frequencies, and the fundamental frequency $\omega_1$ is related to microscopic parameters by
\begin{equation}
    \omega_1=\frac{\pi}{L_x\sqrt{lc}}=\frac{e\pi}{L_x}(\frac{2d n_x}{L_y\epsilon m})^{1/2}.
\end{equation}
The explicit mode decomposition is presented in the Appendix \ref{appendix_mode_expansion}. As an order-of-magnitude estimate, taking an effective surface-current thickness $L_{z,\mathrm{eff}}\sim10\, \mathrm{nm}$--$100\,\mathrm{nm}$ \cite{Gubin2005PRB,tinkham2004introduction},
which represents the current-carrying depth of the near-surface superconducting current, resonator dimensions $L_x,L_y\sim1\,\mathrm{cm},10\,\mathrm{\mu m}$, spacing $d\sim1\,\mathrm{\mu m}$ \cite{wallraff2004Nature,Goppl2008JAP}, and surface effective dielectric constant $\epsilon\sim\times10^{-11}\,\mathrm{F/m}$--$10^{-10}\,\mathrm{F/m}$ \cite{Goppl2008JAP}, together with electron mass $m\sim 9\times10^{-31}\,\mathrm{kg}$ and density $n\sim 10^{28}\,\mathrm{m}^{-3}$ \cite{OpenStaxUP3}, and using $n_x=nL_{z,\mathrm{eff}}L_y$, one obtains the estimate $\omega_1/2\pi\sim10\,\mathrm{GHz}$--$100\,\mathrm{GHz}$, which is close to the typical experimental scale reported for superconducting transmission-line resonators \cite{wallraff2004Nature,Goppl2008JAP}.

Furthermore, to connect with the usual phenomenological transmission-line theory \cite{Blais2004PRA}, we define the nonlocal macroscopic charge operator
\begin{equation}
    \hat Q(x)=-e\int^x_0\mathrm{d}x'\hat n(x'),
\end{equation}
Using Eq.~(\ref{current}), one finds
\begin{equation}
    [\hat Q(x),-l\hat I(x')]=\mathrm{i}\hbar\delta(x-x').
\end{equation}
If $-l\hat I(x)$ is identified as the canonical conjugate to $\hat Q(x)$, the classical Lagrangian corresponding to the Hamiltonian (\ref{effective_Hamiltonian_quantum}) is recovered as
\begin{align}
    \mathcal{L}=&\,\int_0^{L_x}\mathrm{d}x(\frac{l}{2}I^2-\frac{1}{2c}(\partial_xQ)^2)\\
    =&\,\int_0^{L_x}\mathrm{d}x(\frac{1}{2l}\dot Q^2-\frac{1}{2c}(\partial_xQ)^2).
\end{align}
This is precisely the standard Lagrangian of the phenomenological transmission-line theory \cite{Blais2004PRA}.

We therefore conclude that the canonical commutation relation between the macroscopic variables $\hat I(x)$ and $\hat Q(x)$ in transmission-line resonator is not a fundamental postulate. Rather, it emerges naturally from the order parameter phase generated by spontaneous symmetry breaking once that phase is subjected to \textit{third quantization}.
\begin{figure*}
    \centering
    \includegraphics[width=1\textwidth]{macroscopic_variables_correspoding.pdf}
    \caption{Correspondence between representative superconducting circuit elements, the effective Hilbert spaces in which they are described, the commutation relations emerging in \textit{third quantization}, and the associated macroscopic observables.}
    \label{fig3}
\end{figure*}

\section{Third quantization in Quantum circuits}
As summarized in Fig.~\ref{fig3}, and building on the preceding results, we establish a systematic correspondence between the two basic conjugate relations arising in \textit{third quantization} -- namely, the global phase $\hat \phi$ and the total electron number $\hat N$, as well as their local extension, the phase $\varphi(x)$ (i.e., $\phi(x)$) and $n(x)$ -- and the principal macroscopic variables appearing in superconducting quantum circuits. From this perspective, the quantization of superconducting circuit elements in circuit QED can be described within \textit{third quantization}, rather than as a collection of disconnected constructions.

Since superconducting qubit architectures are built from Josephson junction \cite{Review_qubit_computation2001}, we begin by considering a Josephson junction and denote the superconductors on its two sides by $L$ and $R$. When the system is in the ground-state space spanned by $|GS[\Psi(\phi_L)]\rangle\otimes|GS[\Psi(\phi_R)]\rangle$, the Hamiltonian of the junction is \cite{Review_qubit_computation2001,ZLZhang2025PLA}
\begin{equation}
    \hat H_{\mathrm{JC}}=E_C\Delta \hat N^2+E_J\cos\Delta\hat \phi,\label{josephson}
\end{equation}
where $E_J$ is the Josephson energy, $E_C$ is the capacitive energy in the junction, and  $\Delta\hat\phi:=\hat\phi_L-\hat\phi_R$ and $\Delta  \hat N= \hat N_L- \hat N_R$ denote the phase difference and the electron number difference between the two superconductors, respectively. Detailed derivation and discussion about this effective Hamiltonian are given in Appendix \ref{Appendix_Josephson}. The degrees of freedom of this structure therefore originate from the \textit{third quantization} relation between the global order parameter phase and particle number. From this viewpoint, the quantum nature of the macroscopic variables in charge qubits and phase qubits can be understood as different manifestations of the same conjugate relation in different circuit structures.

On the other hand, in the presence of an electromagnetic excitation with $\bm A\neq\bm0$, the system is no longer confined to the ground state space, but instead be described within the low energy excitation space $\{|\varphi(x)\rangle\}$ (equivalently $\{|\phi(x)\rangle\}$). In this setting, the local commutation relation (\ref{commutation_relation_local}) of \textit{third quantization}, provides the origin of the non-commutativity of macroscopic variables in superconducting transmission-line resonators and flux qubits. 

It is worth emphasizing that, the flux variables $\Phi(x)$, which is commonly taken as a field variable in the quantization of superconducting transmission-line resonators \cite{Blais2004PRA,Blais2021RMP}, is in fact an effective flux introduced at the level of the LC circuit description through the relation
\begin{equation}
    \partial_x\Phi(x):=lI.
\end{equation}
In the resonator setting, this quantity does not represent a genuine magnetic flux, i.e., $\oint\bm A\cdot\mathrm{d}\bm l$, rather, it serves more directly as an effective circuit variable encoding the dynamics of the current degree of freedom. Its deeper physical content is therefore rooted not in the effective flux itself, but in the superconducting order parameter phase $\phi(x)$. The \textit{third quantization} developed here starts from the microscopic theory of superconductivity, and takes the phase degree of freedom as the object of quantization, thereby providing a deeper superconducting origin for the phenomenological quantization of transmission-lines.

\section{Discussion and Conclusion}
As a central component of superconducting circuits, the superconducting transmission-line resonator is conventionally quantized by modeling it as an effective LC circuit and imposing canonical commutation relations phenomenologically on macroscopic variables such as charge and flux. From a microscopic perspective, however, this standard procedure leaves two basic questions unresolved: whether these commutation relations should be regarded as fundamental postulates at the macroscopic level, and how the quantization of circuit variables is rooted in the spontaneous $U(1)$ symmetry breaking of the superconducting state.

In this work, starting from the microscopic theory of superconductivity, we have established an intrinsic connection between the quantization of superconducting circuits and the quantization of the superconducting order-parameter phase. We have shown that the order-parameter phase $\phi$, which emerges after spontaneous symmetry breaking in a many-body superconducting system described within second quantization, can be quantized within the ground-state manifold as an operator canonically conjugate to the total Cooper-pair number operator $\hat N_c$, satisfying $[\hat N_c,\hat\phi]=\mathrm{i}$. Moreover, we have demonstrated that this construction extends consistently from the global phase to the spatially local case in the low-energy phase-excitation subspace, where $[\hat n_c(x),\hat\phi(x')]=\mathrm{i}\delta(x-x')$. In this sense, the quantization of the order-parameter phase constitutes what we call \textit{third quantization}. Our results show that \textit{third quantization} is not an additional postulate, but rather a natural structural level that emerges in superconducting systems with spontaneous symmetry breaking.

Going beyond the conventional phenomenological description, we have applied this third-quantization framework to the superconducting transmission-line resonator. Starting directly from the microscopic pairing Hamiltonian underlying BCS superconductivity, we derived its low-energy effective Hamiltonian, including the contributions associated with distributed capacitive and inductive elements, and established quantitative relations between macroscopic observables—such as current and voltage—and the spatially local superconducting phase, as well as the microscopic parameters of the underlying electron-phonon system. This provides a clear microscopic foundation for the non-commutativity of macroscopic circuit variables: it should not be viewed as a newly imposed fundamental principle at the circuit level, but rather as a macroscopic quantum effect that arises from the \textit{third quantization} for the superconducting order parameter in the low-energy effective theory.

On this basis, we further established a systematic correspondence between the macroscopic variables of different superconducting circuit elements and the fundamental conjugate degrees of freedom appearing in \textit{third quantization}. This places transmission-line resonators, capacitive and inductive circuit elements, and related superconducting qubit degrees of freedom within a common conceptual framework. From this viewpoint, the quantization rules used in circuit QED are not independent phenomenological assumptions, but manifestations of a unified microscopic mechanism.

More broadly, our work suggests that the macroscopic quantum properties of a wide class of superconducting circuit elements originate from the same underlying source: the \textit{third quantization} for the superconducting order parameter. This perspective provides a new microscopic foundation for the quantization of superconducting circuits and offers a coherent framework for understanding macroscopic quantum phenomena in more general superconducting architectures.

Finally, we note that both paper (I) and the present paper are restricted to idealized systems. When environmentally induced dissipation is no longer negligible, the macroscopic quantum behavior may be substantially modified \cite{Leggett_AnnPhys1983,CPSun_PRA1994,CPSun_PRA1995}, and \textit{third quantization} itself may require further extension and refinement in the setting of open systems. Addressing these issues will help clarify the conditions under which macroscopic quantum behavior can persist robustly in systems such as superconducting circuits. This remains an important direction for future work.


\funding{This work was supported by the Science Challenge Project (Grant No.TZ2025017), the National Natural Science Foundation of China (NSFC) (Grant Nos. 12088101, 12547124), and the China Postdoctoral Science Foundation (Grant No. 2025M784438).}


\data{Data supporting the findings of this study are available from the authors upon reasonable request.}


\appendix
\section{Real-Space Variational Ground State and Its Self-Consistency Equation}\label{Appendix_variant}
In this Appendix, we present the detailed derivation of the real-space variational ground state in Eq.~(\ref{viriational_ground_state}) and its self-consistency equation in Eq.~(\ref{self_consistency_main_text}) used in Sec. \ref{sec2}. Starting from the attractive Hubbard model, we minimize the free-energy functional and obtain the variational equation for the pairing field $\Psi_{\bm i \bm j}$. We then decompose
$\Psi_{\bm i\bm j}$ into its real amplitude and phase. This gives a uniform-phase solution $\phi_{\bm i\bm j}=\phi$, $\forall \bm i,\bm j$, which corresponds to the spontaneous $U(1)$ symmetry breaking. It also gives the self-consistency equation for the amplitude $|\Psi_{\bm i\bm j}|$.

\subsection{ground state and its self-consistency equation}\label{AppendixA1}
For later convenience, in this Appendix, we we relabel the three-dimensional lattice index $\bm i=(i_x,i_y,i_z)$ used in the main text by a single index $i$. For example, one may take $i:=i_x+N_x(i_y-1)+N_xN_y(i_z-1)$, where $N_x$, $N_y$, $N_z$ are the numbers of lattice sites along the $x-$, $y-$ and $z-$ directions. This relabeling gives a one-to-one correspondence between $i$ and $\bm i$. Therefore, the amplitude $\Psi_{\bm i\bm j}$ in the main text is denoted by $\Psi_{ij}$ in this Appendix, i.e., $\Psi_{ij}\equiv\Psi_{\bm i\bm j}$. Similarly, the lattice operator $\hat c_{\bm i,\sigma}$ in the main text is denoted by $\hat c_{i,\sigma}$ here. This relabeling changes only the notation and does not change the physical content.

In the single-index notation, the attractive Hubbard Hamiltonian in Eq. (\ref{Hubbard_model}) can be written as
\begin{equation}
    H=-t\sum_{\langle ij\rangle,\sigma}\hat c_{i,\sigma}^\dagger \hat c_{j,\sigma}-\sum_{i,\sigma}(eV_{i}+\mu) \hat n_{i,\sigma}-U\sum_{\bm i}\hat n_{i,\uparrow}\hat n_{i,\downarrow}+\mathrm{h.c.}.\label{Hubbard_appendixA}
\end{equation}
The corresponding Gaussian variational ground state is given by
\begin{equation}
    |GS[\Psi]\rangle=\mathcal{N}\exp(\sum_{ij}\Psi_{ij}\hat c_{i,\uparrow}^\dagger \hat c_{j,\downarrow}^\dagger)|0\rangle,\label{viariant_state}
\end{equation}
where $\Psi:=[\Psi_{ij}]$ is the matrix formed by all pairing fields $\Psi_{ij}$. The normalization factor $\mathcal{N}$ satisfies
\begin{equation}
    |\mathcal{N}|^2=|\det\{(1+\Psi^\dagger\Psi)^{-1}\}|.
\end{equation}

The zero-temperature free-energy functional is defined as
\begin{equation}
    F[\Psi]=\langle GS[\Psi]|H|GS[\Psi]\rangle.\label{free_energy_form}
\end{equation}
From Eq. (\ref{Hubbard_appendixA}), we have the functional form in
\begin{equation}
    F[\Psi]=-t\sum_{\langle ij\rangle,\sigma}\langle \hat c^\dagger_{i,\sigma}\hat c_{j,\sigma}\rangle-\sum_{i\sigma}(eV_i+\mu)\langle \hat c_{i,\sigma}^\dagger \hat c_{i,\sigma}\rangle-U\sum_{i}(\langle \hat c_{i,\uparrow}^\dagger \hat c_{i,\uparrow}\rangle\langle    \hat c^\dagger_{i,\downarrow}c_{i,\downarrow}\rangle+|\langle \hat c_{i,\downarrow}\hat c_{i,\uparrow}\rangle|^2),\label{variant_free_energy}
\end{equation}
where $\langle \hat c^\dagger_{i,\sigma}\hat c_{j,\sigma}\rangle:=\langle GS[\Psi]|\hat c^\dagger_{i,\sigma}\hat c_{j,\sigma}|GS[\Psi]\rangle$, and $\langle \hat c_{i,\downarrow}\hat c_{i,\uparrow}\rangle:=\langle GS[\Psi]|\hat c_{i,\downarrow}\hat c_{i,\uparrow}|GS[\Psi]\rangle$ are determined by $\Psi_{ij}$. More specifically, they satisfy
\begin{align}
    \langle \hat c^\dagger_{i\uparrow}\hat c_{j\uparrow}\rangle[\Psi]=&\,\delta_{ij}-[(I+\Psi\Psi^\dagger)^{-1}]_{ji},\label{cicj_Psi1}\\
    \langle \hat c^\dagger_{i\downarrow}\hat c_{j\downarrow}\rangle[\Psi]=&\,\delta_{ij}-[(I+\Psi^\dagger\Psi)^{-1}]_{ij},\label{cicj_Psi2}\\
    \langle \hat c_{i\downarrow}\hat c_{i\uparrow}\rangle[\Psi]=&\,[\Psi(I+\Psi^\dagger\Psi)^{-1}]_{ii}.\label{cici_Psi3}
\end{align}
The detailed derivation of this relation is given in Sec.~\ref{appendixA2}.

We define the average particle number for spin $\sigma$ at site $i$ by
\begin{equation}
    n_{i,\sigma}:=\langle \hat c_{i,\sigma}^\dagger\hat c_{i,\sigma}\rangle.
\end{equation}
Moreover, the superconducting order parameter is defined by
\begin{equation}
    \Delta_i:=-U\langle \hat c_{i,\downarrow}\hat c_{i,\uparrow}\rangle.
\end{equation}
Taking the variation of the free-energy function in Eq.~(\ref{free_energy_form}), we have
\begin{equation}
    \delta F[\Psi]=-t\sum_{\langle ij\rangle,\sigma}\delta\langle \hat c^\dagger_{i,\sigma}\hat c_{j,\sigma}\rangle-\sum_{i\sigma}(eV_i+\mu+Un_{i\bar\sigma})\delta\langle\hat c_{i,\sigma}^\dagger \hat c_{i,\sigma}\rangle+\sum_i\Delta_i\delta\langle \hat c_{i,\uparrow}^\dagger \hat c_{i,\downarrow}^\dagger\rangle+\mathrm{c.c.},\label{variant_equation_1}
\end{equation}
where $\bar\sigma$ denotes the spin opposite to $\sigma$.

Using Eqs.~(\ref{cicj_Psi1}), (\ref{cicj_Psi2}) and (\ref{cici_Psi3}), the variation in Eq.~(\ref{variant_equation_1}) is further written as
\begin{align}
    \delta F[\Psi]=&\,\sum_{\alpha,i,j,l} [(I+\Psi\Psi^\dagger)^{-1}]_{\alpha i}\sum_m\{[-t\sum_{\hat\delta}\delta_{i,m+\hat\delta}-(eV_i+\mu+U n_{i,\downarrow})\delta_{im}]\Psi_{mj}+\notag\\
    &+\Psi_{im}[-t\sum_{\hat\delta}\delta_{j,m+\hat\delta}-(eV_j+\mu+U n_{j,\uparrow})\delta_{mj}]+\Delta_{i}\delta_{im}\delta_{mj}-\label{variant_equation_appendix_amplitude}\\&-\Psi_{im}\Delta_m^*\Psi_{mj}\}[(I+\Psi^\dagger\Psi)^{-1}]_{jl}\delta\Psi^*_{\alpha l}+\mathrm{c.c.}.\notag
\end{align}
The detailed derivation of this equation is given in Appendix \ref{appendixA2}. Since $\Psi_{\alpha l}$ and $\Psi^*_{\alpha l}$ can be treated as independent variables in the variation, we impose $\delta F/\delta\Psi_{\alpha l}=\delta F/\delta\Psi^*_{\alpha l}=0$, $\forall \alpha,j$. This gives the variational constraint equation for the pairing field $\Psi_{ij}$, reads
\begin{equation}
    -t\sum_{\hat\delta}(\Psi_{i+\hat\delta,j}+\Psi_{i,j+\hat\delta})-(eV_i+\mu+U n_{i,\downarrow})\Psi_{ij}-(eV_j+\mu+U n_{j,\uparrow})\Psi_{ij}+\Delta_i\delta_{ij}-\sum_m \Psi_{im}\Delta_m^*\Psi_{mj}=0,\label{constraint_equation_appendix}
\end{equation}
where $i+\hat\delta$ denotes a lattice site adjacent to the site $i$, and $\hat\delta$ runs over the six directions $\pm\hat x$, $\pm\hat y$ and $\pm\hat z$.

We now discuss the phase structure of this constraint equation. We first define $\Psi_{ij}:=|\Psi_{ij}|\exp(\mathrm{i}\phi_{ij})$ and $\Delta_i:=\tilde\Delta_i\exp(\mathrm{i}\chi_{ij})$. Here, $\tilde{\Delta}_i$ is the real amplitude of the order parameter and satisfies
\begin{equation}
    \tilde{\Delta}_i:=-U|\langle\hat c_{i,\downarrow}\hat c_{i,\uparrow}\rangle|=-U[\tilde\Psi(I+\tilde\Psi^\dagger\tilde\Psi)^{-1}]_{ii},
\end{equation}
where $\tilde\Psi:=[|\Psi_{ij}|]$ denotes the real matrix formed by the real amplitudes $|\Psi_{ij}|$.

Substituting these definitions into Eq.~(\ref{constraint_equation_appendix}), and taking the imaginary and real parts separately, we obtain two real equations,
\begin{align}
    &-t\sum_{\hat\delta}[|\Psi_{i+\hat\delta,j}|\sin(\phi_{i+\hat\delta,j})+|\Psi_{i,j+\hat\delta}|\sin(\phi_{i,j+\hat\delta})]-(eV_i+\mu+U n_i^{(\downarrow)})|\Psi_{ij}|\sin(\phi_{ij})\label{image_equation}\\&-(eV_j+\mu+U n_j^{(\uparrow)})|\Psi_{ij}|\sin(\phi_{ij})+\tilde\Delta_i\delta_{ij}\sin(\chi_{i})-\sum_m|\Psi_{im}|\tilde\Delta_m|\Psi_{mj}|\sin(\phi_{im}+\phi_{mj}-\chi_{m})=0,\notag
\end{align}
and
\begin{align}
     &-t\sum_{\hat\delta}[|\Psi_{i+\hat\delta,j}|\cos(\phi_{i+\hat\delta,j})+|\Psi_{i,j+\hat\delta}|\cos(\phi_{i,j+\hat\delta})]-(eV_i+\mu+U n_i^{(\downarrow)})|\Psi_{ij}|\cos(\phi_{ij})\label{real_equation}\\&-(eV_j+\mu+U n_j^{(\uparrow)})|\Psi_{ij}|\cos(\phi_{ij})+\tilde\Delta_i\delta_{ij}\cos(\chi_{i})-\sum_m|\Psi_{im}|\tilde\Delta_m|\Psi_{mj}|\cos(\phi_{im}+\phi_{mj}-\chi_{m})=0.\notag
\end{align}
 These two equations show that the phase $\phi_{ij}$ of the pairing field and the phase $\chi_{ij}$ of the order parameter $\Delta_i$ must be matched with each other. The imaginary part, Eq.~(\ref{image_equation}), determines the phase structure of the variational solution. It gives a solution consistent with the spontaneous breaking of the global $U(1)$ symmetry,
 \begin{equation}
    \phi_{ij}=\chi_{i}=\phi=\mathrm{const},\,\forall\, i,j.
\end{equation}
Under this solution, all real-space pairing components $\Psi_{ij}$ share the same macroscopic phase. The real part, Eq. (\ref{real_equation}), then gives the self-consistency equation for the real pairing amplitude $|\Psi_{ij}|$ is written as
\begin{equation}
    -t\sum_{\hat\delta}(|\Psi_{i+\hat\delta,j}|+|\Psi_{i,j+\hat\delta}|)-(eV_i+\mu+U n_{i,\downarrow})|\Psi_{ij}|-(eV_j+\mu+U n_{j,\uparrow})|\Psi_{ij}|+\tilde\Delta_i\delta_{ij}-\sum_m |\Psi_{im}|\tilde\Delta_m|\Psi_{mj}|=0.
\end{equation}

Restoring the three-dimensional lattice index $i$, we obtain the real-amplitude self-consistency equation used in the main text. This result shows that the spatially inhomogeneous potential changes the distribution of $|\Psi_{ij}|$, while the ground state  $|GS[\Psi(\phi)]\rangle$ can still be labeled by a unified macroscopic phase $\phi$.

\subsection{Relation to the BCS variational state in the uniform limit}\label{BCS}
We now show that, in the spatially uniform limit, the above real-space variational ground state reduces to the usual BCS variational state. The corresponding real-space self-consistency equation also reduces to the BCS self-consistency equation.

We consider a uniform system with $V_i=V$. In this case, the average particle number and the order parameter are then independent of the lattice site. We denote them by $n_{\sigma}:=\sum_{i}\langle\hat c_{i,\sigma}^\dagger \hat c_{i,\sigma}\rangle=n/2$ and $\tilde\Delta_i:=\Delta$, respectively. For convenience, we impose periodic boundary conditions.
Under the conditions, the system is translationally invariant. In particular, the paring amplitude $|\Psi_{ij}|$ depends only on the relative position between two lattice sites,
\begin{equation}
    |\Psi_{ij}|=|\Psi(\bm r_i-\bm r_j)|.
\end{equation}
We then expand it in Fourier components, as in
\begin{equation}
    |\Psi_{ij}|=\frac{1}{N}\sum_{\bm k}\Psi_{\bm k}e^{\mathrm{i}\bm k\cdot(\bm r_i-\bm r_j)},
\end{equation}
furthermore, the momentum-space fermion operator is introduced in
\begin{equation}
    \hat c_{\bm k,\sigma}=\frac{1}{\sqrt N}\sum_i e^{-\mathrm{i}\bm k\cdot\bm r_i}\hat c_{i,\sigma}.
\end{equation}

Using this Fourier transformation, the real-space Gaussian variational ground state $|GS[\Psi(\phi)]\rangle$ is written as
\begin{align}
    |GS[\Psi(\phi)]\rangle=&\,\mathcal{N}\exp(\sum_{\bm k}\Psi_{\bm k}e^{\mathrm{i}\phi}\hat c_{\bm k,\uparrow}^\dagger \hat c^\dagger_{-\bm k,\downarrow})|\bm 0\rangle\\
    =&\,\prod_{\bm k}(\sin\theta_{\bm k}+\cos\theta_{\bm k}e^{\mathrm{i}\phi}\hat c_{\bm k,\uparrow}^\dagger \hat c^\dagger_{-\bm k,\downarrow})|\bm 0\rangle,
\end{align}
where $\cot\theta_{\bm k}:=\Psi_{\bm k}$. This is the standard BCS variational ground state.

We next show that the variational equation reduces to the BCS self-consistency equation in the uniform limit. In the uniform system, the real-space constraint equation can be diagonalized in momentum space. This gives
\begin{equation}
    \frac{1}{N}\sum_{\bm k}[2\xi_{\bm k}\Psi_{\bm k}+\Delta-\Delta\Psi_{\bm k}^2]e^{\mathrm{i}\bm k\cdot(\bm r_i-\bm r_j)}=0,\label{equation_homogeneous}
\end{equation}
where $\xi_{\bm k}$ is the effective energy defined in
\begin{equation}
    \xi_{\bm k}=\epsilon_{\bm k}-(eV+\mu+Un/2).
\end{equation}
The quantity
$\epsilon_{\bm k}=-t\sum_{\hat\delta}\exp(\mathrm{i}\bm k\cdot\hat\delta)=-2t\sum_{a=x,y,z}\cos k_a$ is the kinetic energy.

The solution of Eq.~(\ref{equation_homogeneous}) is given by
\begin{equation}
    \Psi_{\bm k}=-\frac{\Delta}{E_{\bm k}+\xi_{\bm k}},
\end{equation}
where $E_{\bm k}=(\xi_{\bm k}^2+\Delta^2)^{1/2}$. Using this solution, the order parameter satisfies
\begin{equation}
    \Delta=-U\langle\hat c_{i,\downarrow}\hat c_{i,\uparrow}\rangle=-\frac{U}{N}\sum_{\bm k}\langle\hat c_{-\bm k,\downarrow}\hat c_{\bm k,\uparrow}\rangle=-\frac{U}{N}\sum_{\bm k}\frac{\Psi_{\bm k}}{1+\Psi_{\bm k}^2}=\frac{U}{N}\sum_{\bm k}\frac{\Delta}{2E_{\bm k}}.
\end{equation}
Therefore, in the superconducting phase with $\Delta\neq0$, the BCS gap equation is recovered,
\begin{equation}
    1=\frac{U}{N}\sum_{\bm k}\frac{1}{E_{\bm k}}.
\end{equation}

The result shows that when $V_i=V$, both the variational state $|GS[\Psi(\phi)]\rangle$ and the self consistency equation in Eq.~(\ref{self_consistency_main_text}) reduce to the standard BCS forms.

\subsection{Derivation of the relation between $\langle\hat c_{j,\sigma}^\dagger \hat c_{i,\sigma}\rangle$, $\langle\hat c_{j,\downarrow}\hat c_{i,\uparrow}\rangle$, and $\Psi_{ij}$}\label{appendixA2}
In this subsection, we derive the functional relations in Eqs.~(\ref{cicj_Psi1}), (\ref{cicj_Psi2}) and (\ref{cici_Psi3}). For convenience, we first define the matrices $\rho$ and $\kappa$, whose matrix elements are given by
\begin{align}
    \rho^{(\sigma)}_{ij}:=&\,\langle \hat c_{j,\sigma}^\dagger \hat c_{i,\sigma}\rangle,\\
    \kappa_{ij}:=&\,\langle \hat c_{j,\downarrow}\hat c_{i,\uparrow}\rangle.
\end{align}
We also define $\hat B^\dagger$ by
\begin{equation}
    \hat B^\dagger=\sum_{ij}\Psi_{ij}\hat c^\dagger_{i,\uparrow}\hat c^\dagger_{j,\downarrow}.
\end{equation}
Then the variational state is written as $|GS[\Psi]\rangle=\mathcal{N}\exp(\hat B^\dagger)|\bm 0\rangle$. 

Since $\hat c_{i,\sigma}|\bm 0\rangle=0$, we have $\mathcal{N}\exp(B^\dagger)c_{i\sigma}\exp(-B^\dagger)\exp(B^\dagger)|0\rangle=0$ . Thus $\exp(B^\dagger)c_{i\sigma}\exp(-B^\dagger)|GS[\Phi]\rangle=0$. Using the Baker-Campbell-Hausdorff expansion, we have
\begin{align}
    \exp(B^\dagger)\hat c_{i,\uparrow}\exp(-B^\dagger)=&\,\hat c_{i,\uparrow}-\sum_{j}\Psi_{ij}\hat c_{j,\downarrow}^\dagger,\\
    \exp(B^\dagger)\hat c_{i,\downarrow}\exp(-B^\dagger)=&\,\hat c_{i,\downarrow}+\sum_{j}\Psi_{ji}\hat c_{j,\uparrow}^\dagger,
\end{align}
Therefore, the variational state satisfies
\begin{align}
    (\hat c_{i,\uparrow}-\sum_{j}\Psi_{ij}\hat c_{j,\downarrow}^\dagger)|GS[\Psi]\rangle=&\,0,\label{ci_uparrow_B}\\
    (\hat c_{i,\downarrow}+\sum_{j}\Psi_{ji}\hat c_{j,\uparrow}^\dagger)|GS[\Psi]\rangle=&\,0.\label{ci_downarrow_B}
\end{align}

For $\rho_{ij}^{(\uparrow)}=\langle \hat c^\dagger_{j,\uparrow}\hat c_{i,\uparrow}\rangle$, the relation in Eq.~(\ref{ci_uparrow_B}) gives
\begin{equation}
    \rho_{ij}^{(\uparrow)}=\langle \hat c^\dagger_{j,\uparrow}\hat c_{i,\uparrow}\rangle=\sum_{k}\Psi_{jk}^*\langle \hat c_{k,\downarrow}\hat c_{i,\uparrow}\rangle=\sum_{k}\Psi_{jk}^*\kappa_{ik}.\label{rho_kappa}
\end{equation}
On the other hande, Eq.~(\ref{ci_downarrow_B}) gives
\begin{align}
    \kappa_{ik}=&\,\langle c_{k,\downarrow}c_{i,\uparrow}\rangle=-\langle \hat c_{i,\uparrow}\hat c_{k,\downarrow}\rangle\\
    =&\,-\langle \hat c_{i,\uparrow}(-\sum_{m}\Psi_{mk}\hat c_{m,\uparrow}^\dagger|\Psi\rangle)=\sum_{m}\Psi_{mk}\langle\Psi|(\delta_{mi}-\hat c^\dagger_{m,\uparrow}\hat c_{i,\uparrow})|\Psi\rangle\\    
    =&\,\sum_{m}\Psi_{mk}(\delta_{mi}-\rho^{(\uparrow)}_{im}).\label{kappa_rho}
\end{align}
Here $\langle\Psi|\Psi\rangle=1$ is used. Substituting Eq.~(\ref{rho_kappa}) into Eq.~(\ref{kappa_rho}), we obtain
\begin{equation}
    \kappa_{ik}=\Psi_{ik}-\sum_m \Psi_{mk}\sum_{n}\Psi_{mn}^*\kappa_{in}=\Psi_{ik}-\sum_{n}\kappa_{in}(\sum_m\Psi_{mn}^*\Psi_{mk})=\Psi_{ik}-\sum_n\kappa_{in}(\Psi^\dagger \Psi)_{nk},
\end{equation}
therefore,
\begin{equation}
    \sum_{n}\kappa_{in}(\delta_{nk}+(\Psi^\dagger \Psi)_{nk})=(\kappa(I+\Psi^\dagger \Psi))_{ik}=\Psi_{ik}.\label{kappa_Psi_element}
\end{equation}
Hence, in the matrix form, $\kappa$ is given by $\kappa=\Psi(I+\Psi^\dagger \Psi)^{-1}$, Using the identity $\Psi(I+\Psi^\dagger\Psi)^{-1}=(I+\Psi\Psi^\dagger)^{-1}\Psi$, we have
\begin{equation}
    \kappa=\Psi(I+\Psi^\dagger\Psi)^{-1}=(I+\Psi\Psi^\dagger)^{-1}\Psi.\label{kappa_Psi}
\end{equation}

We now return to $\rho^{(\uparrow)}$, from Eq.~(\ref{rho_kappa}), one obtains, in matrix form,
\begin{equation}
    \rho^{(\uparrow)}=\Psi(I+\Psi^\dagger \Psi)^{-1}\Psi^\dagger,\label{rho_uparrow_appendix}
\end{equation}
Using Eq.~ (\ref{kappa_Psi}), this becomes $\rho^{(\uparrow)}=(I+\Psi\Psi^\dagger)^{-1}\Psi\Psi^\dagger$. Therefore
\begin{equation}
    \rho^{(\uparrow)}=I-(I+\Psi\Psi^\dagger)^{-1}.\label{rho_uparrow_Psi}
\end{equation}

$\rho^{(\downarrow)}_{ij}=\langle \hat c^\dagger_{j,\downarrow}\hat c_{i,\downarrow}\rangle$ can be obtained in the same way. From Eq.~(\ref{ci_downarrow_B}), we have
\begin{equation}
    \rho^{(\downarrow)}_{ij}=\langle \hat c^\dagger_{j,\downarrow}\hat c_{i,\downarrow}\rangle=-\sum_{m}\Psi^*_{mj}\langle \hat c_{m,\uparrow} \hat c_{i,\downarrow}\rangle=\sum_{m}\Psi^*_{mj}\kappa_{mi},
\end{equation}
Using Eq.~(\ref{kappa_Psi}), this gives
\begin{equation}
    \rho^{(\downarrow)}=[\Psi^\dagger \Psi(I+\Psi^\dagger \Psi)^{-1}]^T=[I-(I+\Psi^\dagger \Psi)^{-1}]^T.\label{rho_downarrow_Psi}
\end{equation}
 in
Combining the above results in Eqs. (\ref{kappa_Psi}), (\ref{rho_uparrow_Psi}) and (\ref{rho_downarrow_Psi}), we obtain the desired relations between $\langle \hat c^\dagger_{j,\sigma}\hat c_{i,\sigma}\rangle$, $\langle \hat c_{j,\downarrow}\hat c_{i,\uparrow}\rangle$ and $\Psi_{ij}$, reads
\begin{align}
    \langle \hat c^\dagger_{j,\uparrow}\hat c_{i,\uparrow}\rangle[\Psi]=&\,\delta_{ij}-[(I+\Psi\Psi^\dagger)^{-1}]_{ij},\\
    \langle \hat c^\dagger_{j,\downarrow}\hat c_{i,\downarrow}\rangle[\Psi]=&\,\delta_{ij}-[(I+\Psi^\dagger\Psi)^{-1}]_{ji},\\
    \langle \hat c_{j,\downarrow}\hat c_{i,\uparrow}\rangle[\Psi]=&\,[\Psi(I+\Psi^\dagger\Psi)^{-1}]_{ij}.
\end{align}
\subsection{Derivation of the variational equation (\ref{variant_equation_appendix_amplitude})}\label{appendixA3}
Derivation of the variational equation in Eq.~(\ref{variant_equation_appendix_amplitude}). For convenience, we define the matrices $G[\Psi]$ and $K[\Psi]$ by
\begin{align}
    G[\Psi]:=&\,(I+\Psi\Psi^\dagger)^{-1},\\
    K[\Psi]:=&\,(I+\Psi^\dagger\Psi)^{-1}.
\end{align}

Then the variations of $G$ and $K$ are given by
\begin{equation}
    \delta G=-G[(\delta\Psi)\Psi^\dagger+\Psi\delta\Psi^\dagger]G=-[G(\delta\Psi)K\Psi^\dagger+\Psi K(\delta\Psi^\dagger )G],
\end{equation}
and
\begin{equation}
    \delta K=-K[(\delta\Psi^\dagger)\Psi+\Psi^\dagger\delta\Psi]K=-[K(\delta\Psi^\dagger)G\Psi+\Psi^\dagger G(\delta\Psi)K].
\end{equation}
Here we have used the identity
$\Psi K=G\Psi$ (i.e., $\Psi(I+\Psi^\dagger\Psi)^{-1}=(I+\Psi\Psi^\dagger)^{-1}\Psi$).

For $\kappa[\Psi]$, we have
\begin{align}
    \delta\kappa=&\,(\delta G)\Psi+G\delta\Psi=-[G(\delta\Psi)K\Psi^\dagger+\Psi K(\delta\Psi^\dagger )G]\Psi+G\delta\Psi\\
    =&\,G(\delta\Psi)(I-K\Psi^\dagger\Psi)-\Psi K(\delta\Psi^\dagger)G\Psi=G(\delta\Psi)K-\Psi K(\delta\Psi^\dagger)G\Psi.\label{delta_kappa}
\end{align}
Its Hermitian conjugate gives
\begin{equation}
    \delta\kappa^\dagger=K(\delta\Psi^\dagger)G-\Psi^\dagger G(\delta\Psi)K\Psi^\dagger.\label{delta_kappa_dagger}
\end{equation}

Similarly, the variations of $\rho^{(\sigma)}$ are given by
\begin{equation}
    \delta\rho^{(\uparrow)}=-\delta G=G(\delta\Psi)K\Psi^\dagger+\Psi K(\delta\Psi^\dagger )G.\label{delta_rho_uparrow}
\end{equation}
and
\begin{equation}
    [\delta\rho^{(\downarrow)}]^T=-\delta K=K(\delta\Psi^\dagger)G\Psi+\Psi^\dagger G(\delta\Psi)K.\label{delta_rho_downarrow}
\end{equation}

we next use these results to derive Eq.~(\ref{variant_equation_appendix_amplitude}). We first define the matrices $h^{(\sigma)}$ by
\begin{align}
    h_{ij}^{(\uparrow)}=&\,-t\sum_{\hat{\delta}}\delta_{j,i+\hat{\delta}}-(eV_i+\mu+Un_{i,\downarrow})\delta_{ij},\\
    h_{ij}^{(\downarrow)}=&\,-t\sum_{\hat{\delta}}\delta_{j,i+\hat{\delta}}-(eV_i+\mu+Un_{i,\uparrow})\delta_{ij}.
\end{align}
Then Eq.~(\ref{variant_equation_1}) is written as
\begin{equation}
    \delta F[\Psi]=\mathrm{Tr}[h^{(\uparrow)}\delta\rho^{(\uparrow)}+h^{(\downarrow)}\delta\rho^{(\downarrow)}+\bm\Delta^\dagger\delta\kappa+\bm\Delta\delta\kappa^\dagger].\label{variant2}
\end{equation}
where $\bm\Delta:=\mathrm{diag}(\Delta_1,\cdots,\Delta_{N})$. Using Eqs.~(\ref{delta_kappa}), (\ref{delta_kappa_dagger}), (\ref{delta_rho_uparrow}) and (\ref{delta_rho_downarrow}), 
Eq.~(\ref{variant2}) becomes
\begin{align}
    \delta F[\Psi]=&\,\mathrm{Tr}[h^{(\uparrow)}\Psi K(\delta\Psi^\dagger)G+h^{(\downarrow)}[K(\delta\Psi^\dagger)G\Psi]^T-\bm\Delta^\dagger\Psi K(\delta\Psi^\dagger)G\Psi+\bm\Delta K(\delta\Psi^\dagger)G]+\mathrm{c.c.}\\
    =&\,\mathrm{Tr}[G h^{(\uparrow)}\Psi K(\delta\Psi^\dagger)+G\Psi h^{(\downarrow)}{}^TK(\delta\Psi^\dagger)-G\Psi\bm\Delta^\dagger\Psi K(\delta\Psi^\dagger)+G\bm\Delta K(\delta\Psi^\dagger)]+\mathrm{c.c.},\label{delta_F_trace}
\end{align}
Finally, we expand Eq.~(\ref{delta_F_trace}) in matrix elements and obtain
\begin{align}
    \delta F[\Psi]=&\,\sum_{\alpha,i,j,l} G_{\alpha i}\sum_m\{[-t\sum_{\hat\delta}\delta_{i,m+\hat\delta}+(V_i-\mu-U n_{i,\downarrow})\delta_{im}]\Psi_{mj}+\notag\\
    &+\Psi_{im}[-t\sum_{\hat\delta}\delta_{j,m+\hat\delta}+(V_j-\mu-U n_{j.\uparrow})\delta_{mj}]+\Delta_{i}\delta_{im}\delta_{mj}-\\&-\Psi_{im}\Delta_m^*\Psi_{mj}\}K_{jl}\delta\Psi^*_{\alpha l}+\mathrm{c.c.}.\notag
\end{align}
This is the variational expansion used in Appendix \ref{AppendixA1}.

\section{Hermiticity of the Global Phase Operator $\hat\phi$ and the Commutation Relation $[\hat N_c,\hat\phi]=\mathrm{i}$}\label{appendix_phase}

In this Appendix, we give a detailed derivation of the global phase operator and the associated commutation relation with Cooper-pair-number operator. This derivation shows that, even for an inhomogeneous superconductor, a Hermitian macroscopic phase operator $\hat\phi$ can still be defined in the superconducting ground-state space in the thermodynamic limit. The eigen-states of this operator are given by superconducting ground states labeled by different global phases $\phi$. We further show that this phase operator is canonically conjugate to the total Cooper-pair-number operator $\hat N_c:=\hat N/2$, and satisfies the commutation relation $[\hat N_c,\hat\phi]=\mathrm{i}$. Therefore, this Appendix can be viewed as a real-space inhomogeneous extension of the global-phase third-quantization construction developed in our previous work.

We start from the uniform-phase ground-state space ${|GS[\Psi(\phi)]\rangle}$, whose elements are labeled by the global phase $\phi$. To promote $\phi$ to a Hermitian operator describing a physical macroscopic phase degree of freedom, one needs to construct a set of orthonormal phase eigen-states. Since different values of $\phi$ label different superconducting ground states, we will show below that, in the thermodynamic limit, superconducting ground states associated with different phases $\phi$ become mutually orthogonal. They can therefore serve as eigen-states of $\hat\phi$ and be used to define the phase operator $\hat\phi$. To show this, we perform a singular-value decomposition of the pairing matrix $\tilde\Psi:=[|\Psi_{ij}|]$, with matrix elements \begin{equation} |\Psi_{ij}|=\sum_\nu U_{i\nu}\lambda_\nu V_{j\nu}, \end{equation} where $\lambda_\nu\geq0$, and $U:=[U_{i\nu}]$ and $V:=[V_{j\nu}]$ are real unitary matrices. We further define new fermionic modes \begin{align} \hat\alpha_{\nu,\uparrow}^\dagger&\,:=\sum_{i}U_{i\nu}\hat c^\dagger_{i,\uparrow},\\ \hat\alpha_{\nu,\downarrow}&\,:=\sum_{j}V_{j\nu}\hat c_{j,\downarrow}^\dagger, \end{align} with $\{\hat\alpha_{\nu,\uparrow},\hat\alpha_{\nu',\uparrow}^\dagger\}=\delta_{\nu,\nu'}$ and $\{\hat\alpha_{\nu,\downarrow},\hat\alpha_{\nu',\downarrow}^\dagger\}=\delta_{\nu,\nu'}$. The ground state in Eq.~(\ref{ground_state}) becomes \begin{equation} |GS[\Psi(\phi)]\rangle=\prod_{\nu}(\cos\theta_\nu+e^{\mathrm{i}\phi}\sin\theta_\nu\hat\alpha_{\nu,\uparrow}^\dagger\hat\alpha_{\nu,\downarrow}^\dagger)|\bm 0\rangle, \end{equation} where $\tan\theta_\nu:=\lambda_\nu$. When the number $M$ of modes satisfying $\sin\theta_\nu\neq0$ tends to infinity, the overlap \begin{align} &\,\big|\langle GS[\Psi(\phi)]|GS[\Psi(\phi')]\rangle\big|^2\notag\\=&\,\prod_{\nu}^{M\rightarrow\infty}[1-\sin^2(2\theta_\nu)\sin^2(\frac{\phi-\phi'}{2})]\rightarrow\delta_{\phi,\phi'}. \end{align} Therefore, once the number $M$ of pairing modes becomes macroscopic, i.e., in the thermodynamic limit, ground states associated with different global phases become naturally orthogonal. On this basis, we define the continuously normalized phase eigen-states as \begin{equation} |\phi\rangle:=\lim_{\delta\phi\rightarrow0}\frac{1}{\sqrt{\delta\phi}}|GS[\Psi(\phi)]\rangle,\label{eigenvalue_state} \end{equation} which satisfy $\langle\phi|\phi'\rangle=\delta(\phi-\phi')$. The operator $\hat\phi$ is therefore defined as the Hermitian operator \begin{equation} \hat\phi:=\int_0^{2\pi}\phi\mathrm d\phi|\phi\rangle\langle\phi|. \end{equation} Furthermore, from Eqs.~(\ref{ground_state}) and (\ref{eigenvalue_state}), we obtain \begin{equation} -2\mathrm{i}\partial_\phi|\phi\rangle=\hat N|\phi\rangle. \end{equation} Therefore, for an arbitrary state $|\varphi\rangle\in\{|\phi\rangle\}$ in the ground-state space, with $f(\phi):=\langle\phi|\varphi\rangle$, \begin{align} \langle\phi|\hat N|\varphi\rangle=&\,-2\mathrm{i}\int^{2\pi}_0\mathrm{d}\phi' f(\phi')\langle\phi|\partial_\phi|\phi\rangle\\ =&2\mathrm{i}\partial_\phi f(\phi), \end{align} hence the total cooper-pair-number operator $\hat N_c:=\hat N/2$ is represented in the $|\phi\rangle$ basis as $\hat N_c=\mathrm{i}\partial_\phi$. It follows that \begin{equation} [\hat N_c,\hat\phi]=\mathrm{i}.\label{commutation_relation} \end{equation} It is shown that, within the ground state space, the uniform macroscopic order parameter phase $\hat\phi$ and the Cooper-pair-number operator $\hat N_c$ form a canonically conjugate pair.

\section{Canonical Mode Expansion and Eigen-frequencies}\label{appendix_mode_expansion}
In this Appendix, we carry out the canonical mode expansion of the effective Hamiltonian in Eq.~(\ref{effective_Hamiltonian_quantum}). This gives the normal-mode representation of the superconducting transmission-line resonator and relates its eigen-frequencies to the microscopic parameters derived in the main text.

The effective Hamiltonian in Eq.~(\ref{effective_Hamiltonian_quantum}) is
\begin{equation}
    \hat H_{\mathrm{eff}}=\int_0^{L_x} \mathrm{d}x\frac{1}{2l}(\frac{\hbar}{2e}\partial_x\hat\varphi)^2+\int_0^{L_x} \mathrm{d}x\frac{1}{2c}(e\hat n(x))^2 .
    \label{effective_Hamiltonian_quantum_appendix}
\end{equation}
Here $l$ and $c$ are the effective inductance and capacitance per unit length, respectively. The field $\hat\varphi(x)$ is the superconducting order-parameter phase, and $\hat n(x)$ is the electron-number line density. They satisfy
\begin{equation}
    [\hat n(x),\hat\varphi(x')]=2\mathrm i\delta(x-x') .
\end{equation}

From the Heisenberg equations of motion generated by Eq.~(\ref{effective_Hamiltonian_quantum_appendix}), we have
\begin{align}
    \partial_t\hat\varphi(x,t)=&\,-\frac{2e^2}{\hbar c}\hat n(x,t),
    \\
    \partial_t\hat n(x,t)=&\,\frac{\hbar}{2e^2 l}\partial_x^2\hat\varphi(x,t).
\end{align}
Therefore, both $\hat\varphi$ and $\hat n$ satisfy the same wave equation,
\begin{equation}
    lc\,\partial_t^2\hat{\mathcal O}(x,t)-\partial_x^2\hat{\mathcal O}(x,t)=0,\qquad\hat{\mathcal O}=\hat\varphi,\hat n.
\end{equation}

For the structure shown in Fig.~\ref{fig_resonator_sketch}, we impose the boundary condition of vanishing current (\ref{current}) at both ends of the superconducting transmission-line resonator, then
\begin{equation}
    \partial_x\hat\varphi(0,t)=\partial_x\hat\varphi(L_x,t)=0.
\end{equation}
Therefore, $\hat\varphi(x,t)$ and $\hat n(x,t)$ can be expanded as
\begin{align}
    \hat\varphi(x,t)=&\,\frac{2e}{\hbar}\sum_{j=1}^{\infty}(\frac{\hbar\omega_j l}{L_x})^{1/2}\frac{L_x}{j\pi}\cos(\frac{j\pi x}{L_x})[\hat a_j e^{-\mathrm i\omega_j t}+\hat a_j^\dagger e^{\mathrm i\omega_j t}],\notag\\\hat n(x,t)=&\,-\mathrm i\sum_{j=1}^{\infty}(\frac{\hbar\omega_j c}{e^2L_x})^{1/2}\cos(\frac{j\pi x}{L_x})[\hat a_j e^{-\mathrm i\omega_j t}-\hat a_j^\dagger e^{\mathrm i\omega_j t}],
    \label{mode_expansion_phi_n}
\end{align}
where the bosonic operators satisfy $[\hat a_i,\hat a_j^\dagger]=\delta_{ij}$. The eigen-frequencies are
\begin{equation}
    \omega_j=\frac{j\pi}{L_x\sqrt{lc}}=j\omega_1 .
\end{equation}

Substituting the mode expansion in Eq.~(\ref{mode_expansion_phi_n}) into Eq.~(\ref{effective_Hamiltonian_quantum_appendix}), the effective Hamiltonian becomes
\begin{equation}
    \hat H_{\mathrm{eff}}=\sum_{j=1}^{\infty}\hbar\omega_j(\hat a_j^\dagger\hat a_j+\frac{1}{2}).
\end{equation}
The fundamental frequency is therefore
\begin{equation}
    \omega_1=\frac{\pi}{L_x\sqrt{lc}} .
\end{equation}
Using the microscopic expressions for the effective inductance and capacitance per unit length in Eq.~(\ref{inductance_per_unit}) and (\ref{capacitance_per_unit}), we obtain
\begin{equation}
    \omega_1=\frac{e\pi}{L_x}(\frac{2dn_x}{L_y\epsilon m})^{1/2}.
\end{equation}
This expression connects the resonator frequency directly to the microscopic parameters of the superconducting system and to the macroscopic parameters of the transmission-line structure.

\section{Derivation of the Josephson Effective Hamiltonian in Eq.~(\ref{josephson})}\label{Appendix_Josephson}
In this appendix, starting from the microscopic single-electron tunnelling Hamiltonian, we briefly derive the effective Hamiltonian in the ground-state space of two superconductors, $\{|GS[\Psi(\phi_L)]\rangle\otimes|GS[\Psi(\phi_R)\rangle\}$. The effective Hamiltonian reads
\begin{equation}
    \hat H_{\mathrm{JC}}=E_C( \hat N_L-\hat N_R)^2+E_J\cos(\hat\phi_L-\hat\phi_R).
\end{equation}
This derivation will show that the capacitance term comes from the electrostatic energy generated by the charge imbalance between the two sides. The Josephson phase-coupling term arises from second-order processes of single-electron tunnelling.

We consider two weakly coupled superconductors. The Hamiltonian of the system is
\begin{equation}
    \hat H=\hat H_0+\hat H_T+\hat H_U,\label{Hamiltonian_Josephson_appendix}
\end{equation}
where $\hat H_0=\hat H_L+\hat H_R$ is the Hamiltonian of the two uncoupled superconductors. Each side is described by the attractive Hubbard model,
\begin{equation}
    \hat H_l=-t\sum_{\langle i_lj_l\rangle,\sigma}\hat c_{i_l,\sigma}^\dagger \hat c_{j_l,\sigma}-\sum_{i,\sigma}(eV_{i_l}+\mu_l)\hat n_{i,\sigma}-U_l\sum_{\bm i}\hat n_{i_l\uparrow}\hat n_{i_l\downarrow},\qquad l=L,R.
\end{equation}
The single-electron tunnelling Hamiltonian is
\begin{equation}
    \hat H_T=\hat T_++\hat T_-,
\end{equation}
where
\begin{equation}
    \hat T_+:=\sum_{\langle i_L,j_R\rangle}T \hat c_{i_L,\sigma}^\dagger\hat c_{j_R,\sigma},\quad,\hat T_-=(\hat T_+)^\dagger.
\end{equation}
Here $\langle i_Lj_R\rangle$ denotes nearest-neighbor lattice-site pairs across the junction. For simplicity, we only consider nearest-neighbor single-electron tunnelling across the junction and take the tunnelling amplitude $T\in\mathbb{R}$ to be a constant.

The term $H_C$ is the capacitance term. It comes from the electrostatic energy generated by the charge imbalance between the two superconductors. We assume that each superconductor can be approximated as an equipotential good conductor. Hence
\begin{equation}
    \hat H_C=-\frac{1}{2}e(\hat N_R-\hat N_L)(\hat U_R-\hat U_L),
\end{equation}
We further assume that the electrostatic potential is uniform in the transverse $y-$, $z-$ directions, therefore, $\hat U_R-\hat U_L\simeq-e(N_R-N_L)/C$, where $C$ s the effective capacitance. Therefore, the capacitance term can be written as
\begin{equation}
    \hat H_C=E_C(\hat N_R-\hat N_L)^2,
\end{equation}
where $E_C$ is the effective charging-energy coefficient.

We now derive the effective Hamiltonian of Eq.~(\ref{Hamiltonian_Josephson_appendix}) in the ground-state space $\{|GS[\Psi(\phi_L)]\rangle\otimes|GS[\Psi(\phi_R)\rangle\}$. The projection operator onto this space is defined as
\begin{equation}
    \hat P_0=\int\mathrm{d}\phi_L\mathrm{d}\phi_R|\phi_L,\phi_R\rangle\langle\phi_L,\phi_R|,
\end{equation}
where
$|\phi_L,\phi_R\rangle:=\lim_{\delta\phi_L,\delta\phi_R\rightarrow0}|GS[\Psi(\phi_L)]\rangle\otimes|GS[\Psi(\phi_R)\rangle/\sqrt{\delta\phi_L\delta\phi_R}$ denotes the continuously normalized phase state. The complementary subspace is $Q:=1-P_0$. 

Keeping terms up to second order in the tunnelling Hamiltonian, the effective Hamiltonian of Eq.~(\ref{Hamiltonian_Josephson_appendix}) in the ground-state space is
\begin{align}
    \hat H_{\mathrm{JC}}\simeq&\,\hat P_0\hat H\hat P_0-\hat P_0\hat H_T\hat Q\frac{1}{\hat Q\hat H_0\hat Q-E_0}\hat Q\hat H_T\hat P_0\notag\\
    =&\,E_0\hat P_0+E_C(\hat N_R-\hat N_L)^2-\hat P_0\hat H_T\hat Q\frac{1}{\hat Q\hat H_0\hat Q-E_0}\hat Q\hat H_T\hat P_0,
\end{align}
where $E_0$ is the ground-state energy of the two uncoupled superconductors. In the above derivation. Here we keep only the leading capacitance contribution and the leading nonvanishing tunnelling contribution.

For compactness, we define
\begin{equation}
    \hat R_Q=\hat Q\frac{1}{\hat Q\hat H_0\hat Q-E_0}\hat Q,
\end{equation}
Then the second-order tunnelling contribution becomes
\begin{equation}
    \hat H_{\mathrm{eff}}^{(2)}=-\hat P_0\hat H_T\hat R_Q\hat H_T\hat P_0.\label{eff_2_appendix}
\end{equation}
We now calculate the explicit form of this second-order term. Since $|GS[\Psi(\phi)\rangle=\exp(\mathrm{i}\hat N\phi/2)|GS[\Psi(0)]\rangle$, the phase state can be written as
\begin{equation}
    |\phi_L,\phi_R\rangle=\hat U(\phi_L,\phi_R)|0,0\rangle,
\end{equation}
where $\hat U(\phi_L,\phi_R):=\exp(\mathrm{i}\hat N_L\phi_L/2)\otimes\exp(\mathrm{i}\hat N_R\phi_R/2)$. Using the Baker-Campbell-Hausdorff formula, we obtain
\begin{equation}
    \hat U^\dagger\hat T_\pm\hat U=e^{\mp\mathrm{i}(\phi_L-\phi_R)/2}\hat T_\pm.
\end{equation}
On the other hand, $\hat H_0$ separately conserves $\hat N_L$ and $\hat N_R$, hence $\hat U^\dagger \hat H_0\hat U=\hat H_0$. Moreover, $\hat U^\dagger\hat P_0\hat U=\hat P_0$, which gives $\hat U^\dagger\hat Q\hat U=\hat Q$. It follows that $\hat U^\dagger\hat R_Q\hat U=\hat R_Q$. Using these relations and substituting $\hat H_T=\hat T_++\hat T_-$ into Eq.~(\ref{eff_2_appendix}), we find
\begin{align}
    \hat H_{\mathrm{eff}}^{(2)}\simeq&\,-\int^{2\pi}_0\mathrm{d}\phi_L\mathrm{d}\phi_R|\phi_L,\phi_R\rangle\langle\phi_L,\phi_R|\langle0,0|\hat U^\dagger(\hat T_++\hat T_-)\hat R_Q(\hat T_++\hat T_-)\hat U|0,0\rangle\\
    =&\,-\int^{2\pi}_0\mathrm{d}\phi_L\mathrm{d}\phi_R|\phi_L,\phi_R\rangle\langle\phi_L,\phi_R|\{\langle 0,0|\hat T_+\hat R_Q \hat T_+|0,0\rangle e^{-\mathrm{i}(\phi_L-\phi_R)}+\langle 0,0|\hat T_-\hat R_Q \hat T_-|0,0\rangle e^{+\mathrm{i}(\phi_L-\phi_R)}\notag\\
    +&\,\langle 0,0|\hat T_+\hat R_Q \hat T_-|0,0\rangle+\langle 0,0|\hat T_-\hat R_Q \hat T_+|0,0\rangle\},\label{mid_appendix}
\end{align}

Since $T\in\mathbb{R}$, we have $\langle 0,0|\hat T_+\hat R_Q \hat T_+|0,0\rangle=\langle 0,0|\hat T_-\hat R_Q \hat T_-|0,0\rangle$. We then define
\begin{equation}
    E_J=-2\langle 0,0|\hat T_+\hat R_Q \hat T_+|0,0\rangle.
\end{equation}
With this definition, Eq.~(\ref{mid_appendix}) becomes
\begin{equation}
     \hat H_{\mathrm{eff}}^{(2)}=E_J\cos(\hat\phi_L-\hat\phi_R)-2\mathrm{Re}[\langle 0,0|\hat T_+\hat R_Q \hat T_-|0,0\rangle]\hat P_0
\end{equation}
The second term is independent of the phase. It only renormalizes the constant energy in the ground-state space.

Therefore, after dropping phase-independent constant terms, the effective Hamiltonian is
\begin{equation}
    \hat H_{\mathrm{JC}}=E_C(\hat N_L-\hat N_R)^2+E_J\cos(\hat\phi_L-\hat\phi_R).
\end{equation}
This is the effective Hamiltonian shown in Eq.~(\ref{josephson}).

\bibliographystyle{iopart-num}
\bibliography{reference}

\end{document}